\documentclass[twocolumn,pra,aps,superscriptaddress]{revtex4-2}

\usepackage[latin9]{inputenc}
\usepackage{mathtools}
\usepackage{amsbsy}
\usepackage{amssymb}
\usepackage{amstext}
\usepackage{bm}
\usepackage{xcolor}
\definecolor{darkblue}{rgb}{0, 0, 1}
\newcommand{\RN}[1]{%
	\textup{\uppercase\expandafter{\romannumeral#1}}%
}
\usepackage[unicode=true,pdfusetitle,
bookmarks=false,
breaklinks=false,pdfborder={0 0 1},backref=false,colorlinks=true,allcolors=darkblue]
{hyperref}
\hypersetup{
	bookmarksnumbered=false,bookmarksopen=false}
\makeatletter
\newcommand{\ket}[1]{\vert #1 \rangle}
\newcommand{\bra}[1]{\langle #1 \vert}

\newcommand{\fc}[1]{{\color{black} #1}}
\newcommand{\jonzen}[1]{{\color{black} #1}}
\newcommand{\wyp}[1]{{\color{black} #1}}

\@ifundefined{textcolor}{}
{%
	\definecolor{BLACK}{gray}{0}
	\definecolor{WHITE}{gray}{1}
	\definecolor{RED}{rgb}{1,0,0}
	\definecolor{GREEN}{rgb}{0,1,0}
	\definecolor{BLUE}{rgb}{0,0,1}
	\definecolor{CYAN}{cmyk}{1,0,0,0}
	\definecolor{MAGENTA}{cmyk}{0,1,0,0}
	\definecolor{YELLOW}{cmyk}{0,0,1,0}
}

\newcommand{\beq}{\begin{equation}}
\newcommand{\eeq}{\end{equation}}
\newcommand{\beqa}{\begin{eqnarray}}
\newcommand{\eeqa}{\end{eqnarray}}

\usepackage{amsmath}
\usepackage{graphicx}
\usepackage{amssymb}
\usepackage{txfonts,color}
\makeatother

\begin{document}
\title{Pulse-based variational quantum optimization and meta-learning in superconducting circuit}

\author{Yapeng Wang}
\affiliation{Institute for Quantum Science and Technology, Department of Physics, Shanghai University, Shanghai 200444, China}

\author{Yongcheng Ding}
\email{jonzen.ding@gmail.com}
\affiliation{Institute for Quantum Science and Technology, Department of Physics, Shanghai University, Shanghai 200444, China}
\affiliation{Department of Physical Chemistry, University of the Basque Country UPV/EHU, Apartado 644, 48080 Bilbao, Spain}

\author{Francisco Andr\'es C\'ardenas-L\'opez}
\email{f.cardeans.lopez@fz-juelich.de}
\affiliation{Forschungszentrum J\"ulich GmbH, Peter Gr\"unberg Institute, Quantum Control (PGI-8), J\"ulich, Germany}

\author{Xi Chen}
\email{xi.chen@csic.es}
\affiliation{Instituto de Ciencia de Materiales de Madrid (CSIC),
Cantoblanco, E-28049 Madrid, Spain}
\date{\today}

\begin{abstract}	
Solving optimization problems using variational algorithms stands out as a crucial application for noisy intermediate-scale devices. Instead of constructing gate-based quantum computers, our focus centers on designing variational quantum algorithms within the analog paradigm. This involves optimizing parameters that directly control pulses, driving quantum states towards target states without the necessity of compiling a quantum circuit. In this work, we introduce pulse-based variational quantum optimization (PBVQO) as a hardware-level framework. We illustrate the framework by optimizing external fluxes on superconducting quantum interference devices, effectively driving the wave function of this specific quantum architecture to the ground state of an encoded problem Hamiltonian. Given that the performance of variational algorithms heavily relies on appropriate initial parameters, we introduce a global optimizer as a meta-learning technique to tackle a simple problem. The synergy between PBVQO and meta-learning provides an advantage over conventional gate-based variational algorithms.
\end{abstract}

\maketitle

\section{Introduction}
Quantum computing has emerged as a promising field with the potential to revolutionize computational methodologies. Unique properties such as superposition and quantum entanglement in quantum mechanics offer theoretical speedups by embedding problems into quantum systems. With advancements in experimental techniques, we now find ourselves in the noisy intermediate-scale quantum (NISQ) era, where quantum computers demonstrate capabilities for solving complex problems~\cite{preskill2018}. Among the most intriguing applications of quantum computing in this era is quantum optimization~\cite{moll2018}. Broadly, quantum optimization entails utilizing quantum devices to determine the ground state of a Hamiltonian that encodes a classical cost function. It finds its roots in the adiabatic theorem, wherein a system is prepared in the ground state of a simple Hamiltonian and evolved to the problem Hamiltonian, guaranteeing convergence to the problem's ground state if adiabatic criteria are met. This concept gave rise to quantum annealing and the development of quantum annealers within the analog quantum computing paradigm~\cite{finnila1994,hauke2020,crosson2021}, obviating the need for constructing quantum gates. Subsequently, it inspired the quantum approximate optimization algorithm (QAOA)~\cite{farhi2014,zhou2020} as one of the most renowned variational quantum algorithms (VQAs)~\cite{cerezo2021,bharti2022}. Through incooperation with classical optimizers, gate parameters within the quantum circuit are iteratively optimized to minimize a cost function, typically the energy expectation of a Hamiltonian. This showcases the utility of state-of-the-art quantum devices across interdisciplinary fields~\cite{vikstal2020,chandarana2023,brandhofer2023,ding2024}.

Various approaches have been proposed to enhance the performance of QAOA. It operates as a bang-bang control, alternating between evolving the mixing and problem Hamiltonians~\cite{yang2017}. Optimal control theory enables analytical solutions for optimal procedures within bounded operating times, featuring square-pulse-based QAOA on two sides and smooth annealing in between~\cite{brady2021}. This resembles mimicking QAOA in the digital-analog quantum computing paradigm~\cite{parra2020,headley2022}, where quantum gates and specific Hamiltonian evolutions coexist. Another perspective focuses on the circuit level, akin to quantum machine learning~\cite{biamonte2017}, proposing different structures as ansatz to enhance model performance. For instance, the quantum alternating operator ansatz replaces trivial transverse local operators with entangling operators tailored to specific problems~\cite{hadfield2019}. Additionally, adaptive QAOA dynamically selects the mixer from a pre-selected operator pool~\cite{zhu2022}. Moreover, shortcuts-to-adiabaticity has been introduced from quantum control to quantum computing~\cite{hegade2021}, as parameter searching in QAOA is equivalent to optimizing a discretized annealing schedule. Approximate counterdiabatic terms can be easily implemented in quantum circuits to suppress energy excitation, thereby enhancing the performance of variational quantum optimization~\cite{sun2022,chandarana2022,hegade2022}. These recently proposed QAOA variants, combined with classical optimizers, have been benchmarked up to 28 qubits~\cite{xu2024}.

Here, we aim to introduce an alternative approach to variational quantum optimization by concentrating on optimizing parameters that characterize control pulses. The advantages of pulse-based methods over traditional gate-based approaches are becoming increasingly evident. Pulse-based optimization offers faster state preparation, simpler implementation, and greater freedom in navigating the quantum state space. Recently, pulse-based models have been proposed to enhance variational quantum algorithms, such as extending variational quantum eigensolver (VQE) to Ctrl-VQE~\cite{meitei2021,egger2023}, modifying ansatz to PANSATZ~\cite{meirom2023}, and exploring applications in quantum machine learning~\cite{tao2024}. 
Previous related work in quantum optimization has explored analog versions of QAOA and realized variational coherent annealing by introducing auxiliary terms~\cite{barraza2022,barraza2023}. Additionally, a comparison between gate-model and pulse-based models has been conducted on bipotent quantum architectures where quantum gates are not readily available for arbitrary qubits~\cite{ji2023}. Building upon these previous results, we propose pulse-based variational quantum optimization (PBVQO), with the objective of solving quantum optimization problems at the level of control pulses on realistic quantum devices. Furthermore, we recognize that local optimization of variational quantum algorithms can be computationally expensive, requiring additional circuit depth and numerous measurements. By initializing parameters close to global minima, the number of optimization iterations can be significantly reduced. This observation motivates the exploration of meta-learning techniques, whose feasibility has been demonstrated in gate-model QAOA with counterdiabaticity~\cite{chandarana23meta}.

\jonzen{The remainder of the paper is organized as follows: In Sec. \ref{pulse-based optimization}, we introduce the concept of PBVQO, which is a hardware-agnostic framework requiring single-qubit operations and tunable two-body interactions. Then we test our method by combining it with a digital-analog superconducting circuit architecture in Sec. \ref{sec:hardware}. } In Sec. \ref{numerical}, we demonstrate our protocol through numerical experiments aimed at solving the MAX-CUT problem with up to 8 qubits. To tackle optimization challenges, we utilize a combination of BFGS, a quasi-Newtonian optimizer for parameter updates, and genetic algorithms to pre-select suitable parameter initializations. Additionally, in Sec. \ref{discussion}, we present a comparative analysis with conventional gate-based QAOA to highlight the advantages of our approach. Finally, the conclusion and outlook are presented in Sec. \ref{conclusion}. 

\section{Pulse-based optimization}
\label{pulse-based optimization}

We exemplify PBVQO by solving the MAX-CUT problem. At the hardware level, \jonzen{we assume that} the hardware architecture can only directly realize two-body neighboring interaction as an Ising spin chain or an Ising spin ring by connecting the first and last qubit with periodic boundary conditions. To avoid searching for composite pulses to implement effective many-body interactions, we utilize the architecture to solve 2-regular graphs without any disconnected cycles, meaning the degree of each node is 2, which can be easily embedded into the architecture by topology. Thus, the problem is equivalent to finding the ground state of an antiferromagnetic Ising ring. \jonzen{We assume that the effective hardware Hamiltonian includes only local spin operator and neighbouring interaction, which can be expressed as:
\begin{equation}
\label{eq:PBQO}
H_{\textbf{PBVQO}} = \sum_{j}\frac{\omega_j}{2}\sigma_j^z + P(t)\sigma_j^y\sigma_{j+1}^y,
\end{equation}
where $P(t)$ is the pulse to be optimized,} \wyp{and $\sigma_j^y\sigma_{j+1}^y$ can be replaced by other interaction terms depending on the capabilities of the platform. }\jonzen{This effective Hamiltonian is feasible in various quantum devices without loss of generality. In principle, the pulse ansatz $P(t)$ can be of any form, as long as it can be implemented on the specefic quantum device. }\wyp{However, in practical implementation, not all pulses are realizable, so a filter $F[P(t)]$ is required for pulse.} \jonzen{For simplicity, we illustrate the PBVQO framework by assuming}
that the optimal pulse can be expressed as a series of trigonometric functions
	\begin{equation}
		\label{eq:P}
		P(t)=\sum_{i=1}^{n} A_i \sin\bigg[(2i-1)\pi t+\phi_i\bigg],
	\end{equation}
where $A_i$ are the pulse amplitudes, $t\in[0,T]$ is the operation time, and $\phi_i$ are the phases. \jonzen{The trigonometric functions resemble the annealing schedule used in variational quantum annealing~\cite{unsal2022}, though the mechanism is not exactly the same. In this context, the series allows for the expression of an arbitrary function by providing sufficient freedom with amplitudes and phases.}

On the contrary, the conventional QAOA for MAX-CUT problems prepares the system in the product state $|+\rangle^{\otimes N}$, which is an eigenstate of the mixing Hamiltonian (mixer) $H_{\text{mix}} = \sum_j\sigma_j^x$, and then let the system evolve by alternate application of the problem Hamiltonian $H_p = \sum_{\langle i,j\rangle\in \mathcal{G}}\sigma_i^z\sigma_j^z$ ($\mathcal{G}$ is the graph to be solved) with the mixing ones for $p$ rounds. In other words, we compute the evolution of the system through the product $|\Psi_f\rangle=\prod_j^p\exp(-i\beta_jH_{\text{mix}})\exp(-i\gamma_jH_p)|+\rangle^{\otimes N}$. QAOA aims for finding the set of parameters $\{\beta_j,\gamma_j\}$ that minimize the averaged energy of the problem Hamiltonian, $\langle\Psi_f|H_p|\Psi_f\rangle$ so that $|\Psi_f\rangle$ be the ground state of the problem Hamiltonian. Measuring the $Z$-Pauli strings of the problem, we obtain the max-cut of the graph, if the optimization procedure was successful. From the annealing perspective, finding the parameters $\{\beta_j,\gamma_j\}$ is equivalent to find the optimal schedule function for a digitalized quantum annealing process

\begin{equation}
H_{\textbf{QA}}= B(t) \sum_j\sigma_j^x + \Gamma(t) \sum_{\langle i,j\rangle\in \mathcal{G}}\sigma_i^z\sigma_j^z,
\end{equation}
where $\beta_i^* = B(i\delta t)\delta t$ and $\gamma_i^* = \Gamma(i\delta t)\delta t$ are the optimal parameters describing the schedule.

\jonzen{We realize that in PBVQO, it is not necessary to prepare the system in the state $|+\rangle^{\otimes N}$.} Instead, it is more advantageous to start in the state $|0\rangle^{\otimes N}$, which is an eigenstate of the mixing Hamiltonian in our architecture, $\sum_{j}{(\omega_j}/{2})\sigma_j^z$. Unlike the typical QAOA realization, we do not switch-off the mixing Hamiltonian. Moreover, we  will consider the two-body interactions to be $\sum_{\langle i,j\rangle}\sigma^{y}_i\sigma^{y}_j$, which remain always turned-on. Consquentely, we need to engineer the pulse sequences such that both Hamiltonian encodes the ground state of the MAX-CUT Hamiltonian $\tilde{H}_{p}=\sum_{\langle i,j\rangle\in \mathcal{G}}\sigma_i^x\sigma_j^x$.

The workflow of PBVQO consists of the following steps: We initialize the pulse ansatz by randomly selecting both amplitudes and phases. Then, the pulse is filtered to obtain $F[P(t)]$.
And calculate the dynamics of the mixing and problem Hamiltonian, resulting in the state $|\Psi_f\rangle$. Afterwards, we compute the average of the energy $\langle\Psi_f|\tilde{H}_p|\Psi_f\rangle$ and we update the parameters $\{A_i,\phi_i\}$ using a classical optimizer, such as BFGS. When the optimization converges according to our hyperparameters, we measure the system on the $x$-axis, obtaining the classical configuration of the MAX-CUT problem. We evaluate the PBVQO performance by computing the error rate $R$, defined as
\begin{equation}
R=\biggm|\frac{\langle\Psi_f^*|\tilde{H}_p|\Psi_f^*\rangle-E_g}{E_g}\biggm|,
\label{error}
\end{equation}
where $E_g$ is the ground energy of $\tilde{H}_p$ obtained by numerically diagonalizing the Hamiltonian.

\section{Hardware architecture}
\label{sec:hardware}

\begin{figure}[t]
\includegraphics[width=\linewidth]{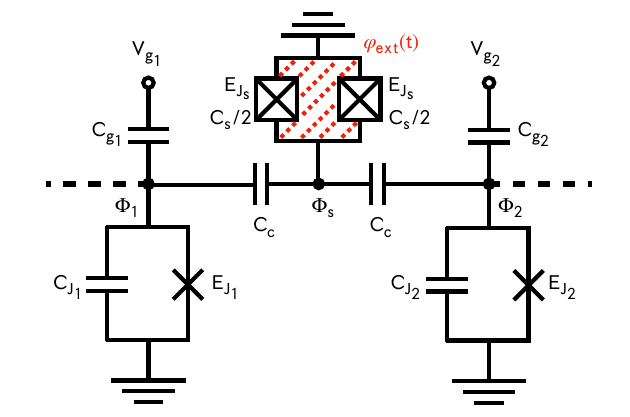}
\caption{Schematic circuit representation of the model implementing the PBVQO. Two charge qubits are coupled through a grounded SQUID. The only tunable parameter is the external flux $\varphi_{\text{ext}}$ through the SQUID, indicated by the red shaded region.}\label{fig:archi} 
\end{figure}

We hereby introduce the superconducting circuit architecture employed for variational quantum optimization, which was originally proposed for digital-analog quantum simulation~\cite{yu2022}. The minimal structure consists of two charge qubits and a grounded superconducting quantum interference device (SQUID) as a coupler. Each charge qubit consists of a capacitance $C_{J\ell}$ parallel connected to a Josephson junction with energy $E_{J\ell}$ and biased with an external voltage source $V_{g\ell}$ as depicted in Fig.~\ref{fig:archi}. Moreover, the SQUID is fromed a closed loop embedded with two identical Josephson junctions of capacitances $C_{s}/2$ and Josephson energies $E_{Js}/2$, respectively. Also, the loop is threaded by an external magnetic flux $\varphi_{ext}(t)$. We denote $\Phi_1$, $\Phi_2$, and $\Phi_s$ as the fluxes describing the charge qubits and the SQUID respectively. The circuit Lagrangian reads
\begin{eqnarray}
\label{eq:Lcircuit}\nonumber
\mathcal{L} &=& \sum_{j=1,2,s}\bigg[\frac{C_{J_{j}}\dot{\Phi}^2_{j}}{2}+\frac{C_{g_{j}}(\dot{\Phi}_{j}-V_{g_{j}})^2}{2}+\frac{C_c(\dot{\Phi}_{j}-\dot{\Phi}_{s})^2}{2}\\
&+&E_{J_{j}}\cos(\varphi_j)\bigg]+\frac{C_{J_{s}}\dot{\Phi}^2_{s}}{2} + E_{J_{s}}^{\rm{eff}}[\varphi_{ext}(t)]\cos(\varphi_s),
\end{eqnarray}
here the overdot notation represents derivation with respect to time. Moreover, $E_{J_{s}}^{\rm{eff}}[\varphi_{ext}(t)]=2E_{J_{s}}|\cos[\varphi_{ext}(t)]|$ is the tunable Josephson energy of the SQUID with $\varphi_{j}=\Phi_{j}/\varphi_{0}$ being the superconducting phase.\par

The idea of this architecture is to manipulate the coupling strength between both charge qubit by modulating the external phase $\varphi_{ext}(t)$ without exciting the SQUID, in this case we need to assume that the latter operates in the so-called \textit{high-plasma frequency} and \textit{low-impedance} approximation~\cite{Nori.2010,Delsing.2011} where it is possible to factorize the SQUID from the system dynamics similar to perturbative diagonalization based on \fc{Schrieffer-Wolff transformation~\cite{bravyi2011}}. Then for eliminating the SQUID degree of freedom, we compute the Euler-Lagrange equation of $\mathcal{L}$ obtaining
\begin{subequations}
\begin{eqnarray}
&&C_{\Sigma_{1}}\ddot{\Phi}_{1}-C_c\ddot{\Phi}_{s}+\frac{E_{J_{1}}}{\varphi_{0}} \sin(\varphi_1)=0,\\
&&C_{\Sigma_{2}}\ddot{\Phi}_{2}-C_c\ddot{\Phi}_{s}+\frac{E_{J_{2}}}{\varphi_{0}} \sin(\varphi_2)=0,\\
&&C_{\Sigma_{s}}\ddot{\Phi}_{s}-C_c(\ddot{\Phi}_{1}+\ddot{\Phi}_{2})+\frac{E_{J_{s}}^{\rm{eff}}[\varphi_{ext}(t)]}{\varphi_{0}} \sin(\varphi_s)=0.
\end{eqnarray}
\end{subequations}
Here, $C_{\Sigma_{j}}$ is the total capacitance of each nodes. The high-plasma frequency allows us to assume that the variation on the fluxes on the charge qubits are larger than in the SQUID i.e., $\ddot{\Phi}_{1(2)}\gg\ddot{\Phi}_{s}$, moreover, the low-impedance regime refers to almost no current flows on the SQUID such that we can operate the SQUID \fc{in the linearized regime~\cite{Nori.2010}}. These conditions permit us to express $\varphi_{s}$ in terms of the other node variables.
\begin{eqnarray}
\label{eq:flux}
{\varphi_s}&=&\sum_{j}{-\frac{C_cE_{J_j}\sin{(\varphi_j)}}{E_{J_s}^{\textrm{eff}}C_{\Sigma_{j}}}}.
\end{eqnarray}
Now, we have to perform the same reasoning to the SQUID charge. To do so, we compute the canonical conjugate momenta $\vec{Q}=\vec{\nabla}_{\vec{\Phi}}\mathcal{L}$
\begin{eqnarray}
\vec{Q}=\begin{pmatrix}
    C_{\Sigma_{1}}\dot{\Phi}_{1}-C_c\dot{\Phi}_{s}-C_{g_1}V_{g_{1}}\\
    C_{\Sigma_{2}}\dot{\Phi}_{2}-C_c\dot{\Phi}_{s}-C_{g_2}V_{g_{2}}\\
    C_{\Sigma_{s}}\dot{\Phi}_{s}-C_c(\dot{\Phi}_{1}+\dot{\Phi}_{2})
\end{pmatrix},
\end{eqnarray}
and apply the high-plasma approximation for eliminating the terms proportional to $\dot{\Phi}_{s}$, leading to
\begin{eqnarray}
\label{eq:charge}
Q_s&=&-C_{c}\sum_j\frac{Q_{j}+C_{g_j}V_{g_j}}{C_{\Sigma_{j}}}.
\end{eqnarray}
The conditions in Eq.~(\ref{eq:flux}) and in Eq.~(\ref{eq:charge}) eliminated the SQUID degree of freedom such that the effective qubit-qubit Hamiltonian can be expressed as
\begin{equation}
\label{eq:Hcircuit2}
H = \sum_{j=1,2}\big[\mathcal{H}_j + \mathcal{H}_{{\rm{ctrl},j}}\big] + \mathcal{H}_{\text{c}},
\end{equation}
where $\mathcal{H}_j$ and are the effective charge qubit and control Hamiltonian, whereas $\mathcal{H}_{\text{c}}$ corresponds to the interaction Hamiltonian given by
\begin{subequations}
\begin{eqnarray}
\mathcal{H}_j &=& \frac{Q_j^2}{2\tilde{C}_{J_j}}-E_{J_j}\cos{\varphi_j} + g[\varphi_{ext}(t)]\sin^{2}{(\varphi_j)},~~~~\\
\mathcal{H}_{{\rm{ctrl},j}} &=& \tilde{n}_{g_{j}}(t)Q_j,~~~~\\
\mathcal{H}_{\text{c}} &=& g[\varphi_{ext}(t)]\sin{(\varphi_1)}\sin{(\varphi_2)}.
\end{eqnarray}
\end{subequations}
where we define the effective capacitaces $\tilde{C}_{J_j}$ and $\tilde{C}_{J_s}$ as
\begin{eqnarray}
\tilde{C}_{J_{j}} &=& \frac{C_*^3}{C_{k}(2C_c+C_s)+C_c(C_c+C_s )},\\
\quad\tilde{C}_{J_s} &=&\frac{C_cC_*^3}{(C_c+C_1)(C_c+C_2)},
\end{eqnarray}
where $C_*^3=C_c(C_1+C_2)(C_s+C_c)+C_c^2C_s+C_1C_2(2C_c+C_s)$, and $C_{j}=C_{g_j}+C_{J_j}$ ($j\in\{1,2\}$ and $k$ is the other element). The dimensionless gate charges are given by
\begin{eqnarray}
\tilde{n}_{g_{j}}(t) &=& -\frac{C_{g_{j}}}{2e\tilde{C}_{J_{j}}}V_{g_{j}}-\frac{\tilde{C}_{J_{j}}C_c^2C_{g_{k}}}{2eC_*^3\tilde{C}_{J_{j}}}V_{g_{k}}.
\label{Eq05}
\end{eqnarray}
Finally, $g[\varphi_{ext}(t)]$ is the effective coupling strength defined as
\begin{equation}
\label{eq:gamma12}
g[\varphi_{ext}(t)]=\frac{C^2_cE_{J_1}E_{J_2}}{E_{J_s}^{\textrm{eff}}(C_1+C_c)(C_2+C_c)}. 
\end{equation}
We quantize the Hamiltonian by promoting quantum operators $\hat{Q}_j=-2e\hat{n}_j$ and $\hat{\varphi}_j$ 
satisfying canonical commutation relation 
$[\hat{n}_j, \exp( \pm i \hat{\varphi}_j)]=\exp(\pm i\hat{\varphi}_j)$, 
substituting Eq.~\eqref{eq:charge} and~\eqref{eq:flux} into the Hamiltonian~\eqref{eq:Hcircuit2}, we obtain for the charge qubit Hamiltonian
\begin{eqnarray}
\mathcal{H}_j = 4E_{Cj}n_j^2-E_{J_j}\cos{\varphi_j} + g[\varphi_{ext}(t)]\sin^{2}{(\varphi_j)}, ~~~
\end{eqnarray}
where $E_{Cj}=e^2/2\tilde{C}_{J_j}$ is the charge energy. In the charge basis representation, the Hamiltonian in Eq.~(\ref{eq:Hcircuit2}) can be expressed in the two-level approximation as follows
\small{
\begin{eqnarray}
\label{eq:tls}
H =\sum_{j=1,2}\bigg[\frac{\omega_j}{2}\sigma_j^z+\Omega_j(t)(n_{0,1}\ket{g}\bra{e}+\text{c.c.})\bigg]+\frac{g[\varphi_{ext}(t)]}{4}\sigma_1^y\sigma_2^y,
\end{eqnarray}}
where $\omega_j=E_{J_j}$ is the transition frequency of the $j$th charge qubit. The effective circuit Hamiltonian has three different contributions; the free energy term setting the transition frequency, the single-qubit interaction with drive amplitude $\Omega_j(t)$ and the tunable interaction term, note as the matrix element of the charge operator are real, we can implement an X or Y rotation just by changing the relative phase of the external charge offset $\tilde{n}_{g_{j}}(t)$. Additionally, in this architecture, depending on the scheduling of the external voltages $V_{g_{j}}(t)$ and the external flux $\varphi_{ext}(t)$ we are able to dynamically \textit{switch off} the two-qubit interaction while implementing the single-qubit rotations and vice versa, allowing to implement the digital-analog QAOA dynamics.\par

On the other hand, the superconducting architecture discussed here is also applicable for implementing gate models by selecting drive frequencies that activate/deactivate certain process in the Hamiltonian in Eq.~(\ref{eq:tls}). This can be achieved by assuming that the external flux contains a constant DC and time varying AC component of the form
\begin{equation}
\varphi_{\text{ext}}(t)=\varphi_{DC}+\varphi_{AC}(t),
\end{equation}
here, we assume that $\varphi_{AC}(t)$ can be decomposed using the pulse defined in Eq.~(\ref{eq:P})
\begin{equation}
\varphi_{AC}(t)=A_1\cos{(\nu_1 t +\tilde{\phi}_1)}+A_2\cos{(\nu_2 t +\tilde{\phi}_2)},
\end{equation}
here, we impose the condition $|A_{1}|, |A_{2}|\ll |\varphi_{DC}|$ such that the effective inductance can be expressed as
\begin{eqnarray}
\frac{1}{E_{J_s}^{\textrm{eff}}} &\approx& \frac{1}{\bar{E}_{J_s}}\left[1+\frac{\sin{({\varphi}_{DC})}}{\cos{({\varphi}_{DC}})}{\varphi}_{AC}(t)\right],\nonumber\\
\bar{E}_{J_s} &=& 2E_{J_s}\left|\cos{({\varphi}_{DC})}\right|.
\end{eqnarray}
Therefore, we reformulate the Hamiltonian~\eqref{eq:tls} as
\begin{equation}
H=\sum_{j=1,2}\bigg[\frac{\omega_j}{2}\sigma_j^z+\frac{\Omega_j(t)}{2}\sigma_j^x\bigg]+\left[g_0+g_1\varphi_{AC}(t)\right]\sigma_1^y\sigma_2^y,
\end{equation}
where the always-on and tunable coupling strength reads
\begin{eqnarray}
g_0&=&\frac{C_{c}^2E_{J_1}E_{J_2}}{4(C_{1}+C_c)(C_{2}+C_c)\bar{E}_{J_s}},\nonumber\\
g_1&=&\frac{C_{c}^2E_{J_1}E_{J_2}}{4(C_{1}+C_c)(C_{2}+C_c)\bar{E}_{J_s}}\frac{\sin{({\varphi}_{DC})}}{\cos{({\varphi}_{DC})}} \, .
\end{eqnarray}
We now proceed by expressing the Hamiltonian in the interaction picture, assumign that there is no drive $\Omega_j(t)=0$, and performing the rotating wave approximation obtaining the Hamiltonian
\begin{align}
\hat{\mathcal{H}}_I =\frac{g_1}{4}\left(M_{-}\sigma_1^x\sigma_2^x+M_{+}\sigma_1^y\sigma_2^y
-N_{+}\sigma_1^x\sigma_2^y
+N_{-}\sigma_1^y\sigma_2^x\right),
\label{Eq21}
\end{align}
where $M_{\pm} = A_1\cos{\tilde{\phi}_1}\pm A_2\cos{\tilde{\phi}_2}$ and $N_{\pm} =A_1\sin{\tilde{\phi}_1} \pm A_2\sin{\tilde{\phi}_2} $.
This approach allows us to engineer the effective two-body interaction required in the algorithm by the adequate selection of the amplitudes and phases of the external flux $\varphi_{AC}(t)$. For instances, we can engineer the standard QAOA the mixing Hamiltonian $XX$ by setting $\tilde{\phi}_1=\tilde{\phi}_2=0$ and $A_1=-A_2$ obtaining 
\begin{equation}
H_{XX} = \frac{g_1A_1}{2}\sigma_1^x\sigma_2^x=G\frac{A_1}{2\cos(\varphi_{DC})}\frac{\sin(\varphi_{DC})}{\cos(\varphi_{DC})}\sigma_1^x\sigma_2^x.
\end{equation}
Thus, this architecture allows us to implement directly the problem-Hamiltonian dynamics $U(-i\beta\sigma_1^x\sigma_2^x)$, making the approach more efficient than compile the problem in terms of single-qubit rotations and entangling CNOT gates. The price to pay is that the condition $|A_1|\ll|\varphi_{DC}|$ bounds our maximal gating time, making the algorithm susceptible to noise and decoherence. A naive approach for accelerating the gating time relies upon setting DC component close to $\varphi_{DC}=\pi/2$. Nevertheless, such approach is not valid because the energy spectrum of the system shrinks, leading to divergencies~\cite{Deshpande2022}. Another approach is to use the architecture for engineering the counter-diabatic (CD) Hamiltonian of the QAOA~\cite{chandarana2022,sun2022}: $XY+YX$ for other selection of the pulse parameters. Thus, the scheduling of the whole QAOA may include the pulses for the single-qubit rotations, the action of the entangling gates and the inclusion of the CD evolution. \par
\fc{Notice that from the Hamiltonian perspective, we can implement PBVQO by setting $P(t)=g[\varphi_{ext}(t)]/4$. However, the proposed trigonometrical ansatz does not match such coupling because $g[\varphi_{ext}(t)]$ is positive. Moreover, we may encounter errors due to the always-on coupling $g_0$ during PBVQO. To address these issues, we can filter our pulses considering such experimental constrains under the relation:
	\begin{eqnarray}
		\label{eqn:F}
		F[P(t)] = \begin{cases}
			~G, & ~~ -G\leq P(t)<G\\
			|P(t)|, & ~~ \text{otherwise}
		\end{cases},
	\end{eqnarray}
where $G={C^2_cE_{J_1}E_{J_2}}/{8E_{J_s}(C_1+C_c)(C_2+C_c)}$ is the bound on the coupling strength.}  

\section{Numerical experiments}
\label{numerical}

\subsection{Baseline}
PBVQO imposes no limit on the operation time; thus, control pulses can be periodic in time, mimicking Floquet engineering protocols. In light of this, our first step is to analyze the role of the pulse duration $T$ in our model's performances, making it a free parameter in our optimization. We employ the BFGS algorithm as the classical optimizer, a quasi-Newton method for solving unconstrained nonlinear optimization problems. It achieves gradient descent by preconditioning the gradient without explicitly requiring the Hessian matrix.

\begin{figure}
\includegraphics[width=1\linewidth]{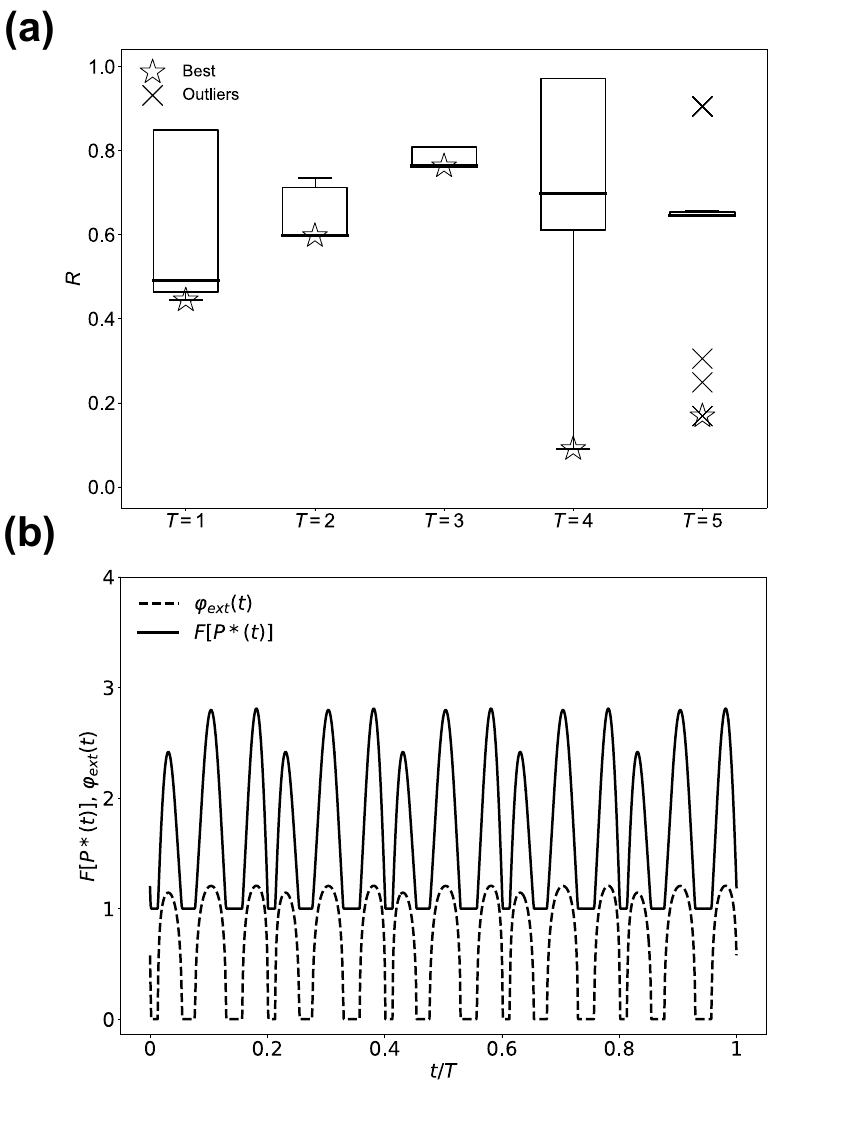}
\caption{(a) Error rate $R=|({\langle\Psi_f^*|\tilde{H}_p|\Psi_f^*\rangle-E_g})/{E_g}|$ of PBVQO for 2-regular graph MAX-CUT problem with 8 qubits. The boxplot represents the inter-quartile range (IQR) of the data, with the box spanning from the first quartile to the third quartile, and a line indicating the median. Whiskers extend from the box to the farthest data points within 1.5 times the IQR. Any data point beyond the whiskers is considered an outlier, denoted by a cross in the figure. Each best model performance over $50$ random initializations of different operation times $T$ is denoted by a star. (b) Optimal variational pulses $F[P^*(t)]$ and corresponding external flux $\varphi_{\text{ext}}(t)$ at $T=5$. It results in the best error rate of $R=0.168$. Optimal parameters: $\{A_1^*= 2.017,,~A_2^*=0.644,~A_3^*=1.384,~\phi_1^*=-0.141,,~\phi_2^*=-0.596,~\phi_3^*=-0.408\}$. Dimensionless parameters for simulation of quantum dynamics: $\omega_1=\omega_2\ = \cdots=\omega_8=6$ and $G=1$.}\label{fig:bfgs_only} 
\end{figure}
In Fig.~\ref{fig:bfgs_only}(a), we solve the MAX-CUT problem for a 2-regular graph using $50$ different randomized initial parameters in each set, with $n=3$ and 6 tunable parameters in the control pulses~\eqref{eqn:F}. To illustrate the dispersion and skewness of the data, we depict the distribution using box plots. We observe that extending the pulse length does not improve model performance; specifically, $T=1$ is sufficient for the problem, although the performance is not entirely satisfactory. The BFGS algorithm often converges at a local minimum, hindering optimal model performance and preventing the attainment of the optimal solution of MAX-CUT after obtaining the classical string via measurement on the $x$-axis. In Fig.~\ref{fig:bfgs_only}(b), we plot the control pulse and corresponding external flux for the best performance among all the initializations, which achieves an error rate of $R=0.168$.

\begin{figure}
\includegraphics[width=1\linewidth]{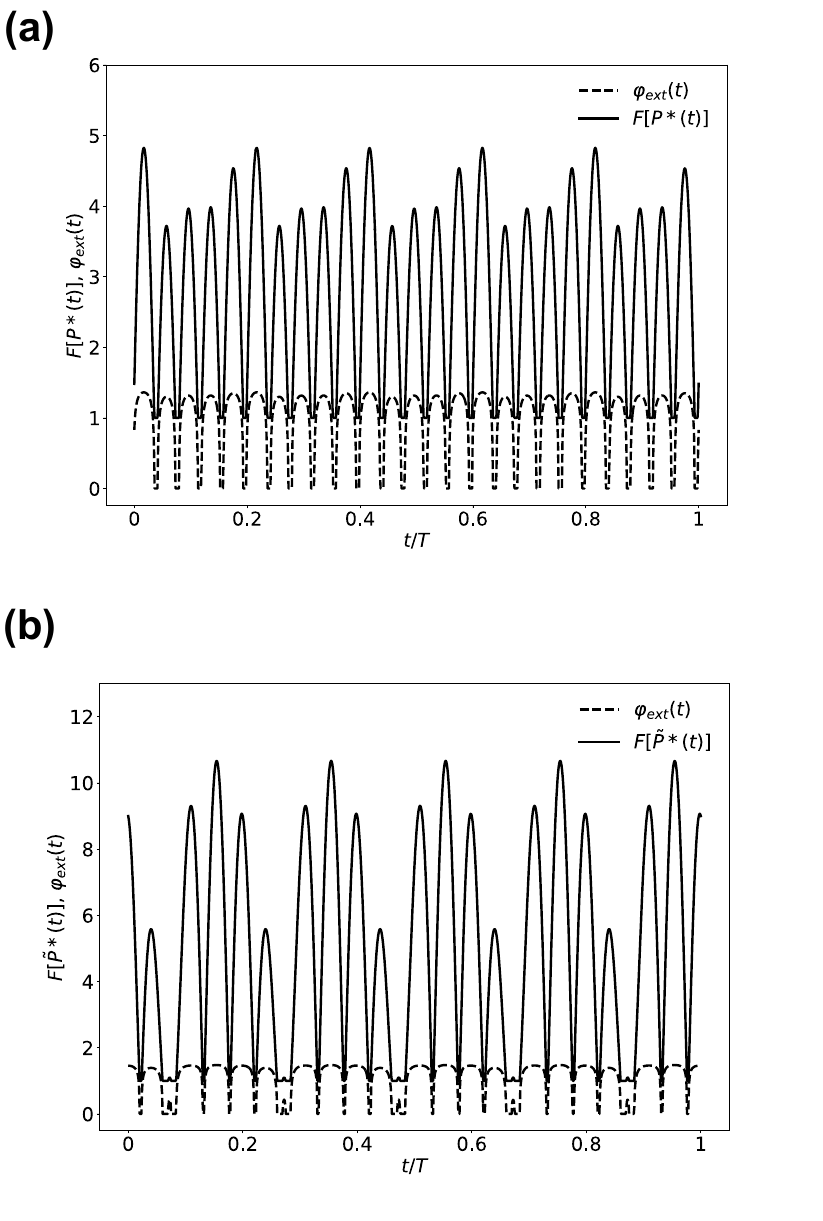}
\caption{(a) Optimal variational pulses $F[P^*(t)]$ and corresponding external flux $\varphi_{\text{ext}}(t)$ found by genetic algorithm for 2-qubit problem. Optimal parameters: $\{A_1^*= 0.307,~A_2^*=0.491,~A_3^*=4.202,~\phi_1^*=3.798,~\phi_2^*=3.253~\phi_3^*=3.441\}$. (b) Optimal variational pulses $F[\tilde{P}^*(t)]$ and corresponding external flux $\varphi_{\text{ext}}(t)$ optimized by BFGS for 8-qubit problem, which is initialized by $\{A_i^*,\phi_i^*\}$ for 2-qubit problem with maximal fitness. Optimal Parameters: $\{\tilde{A}_1^*= -1.668,~\tilde{A}_2^*=4.560,~\tilde{A}_3^*=6.861,~\tilde{\phi}_1^*=3.456,~\tilde{\phi}_2^*=3.919~\tilde{\phi}_3^*=5.113\}$. Dimensionless parameters for simulation of quantum dynamics are the same as those before.}\label{fig:pulse_meta}  
\end{figure}

\subsection{Meta-learning}
As we can see, the performance of variational algorithms heavily relies on parameter initialization. Adequate initialization can bring parameters close to the global minima, significantly reducing the iteration number. In the case of QAOA, one can utilize $\beta_i = B(i\delta t)\delta t$ and $\gamma_i = \Gamma(i\delta t)\delta t$ as digitized annealing schedules for $B(t)$ and $\Gamma(t)$, for instance, $B(t)=1-t/T$ and $\Gamma(t)=t/T$, instead of random initial parameters. However, the tunable pulse $P(t)$ does not conform to the standard quantum annealing paradigm since it is periodic, and the mixer is never turned off. In other words, there is no established initialization strategy for PBVQO. We propose a hypothesis that the optimal parameters $\{A_i^*,\phi_i^*\}$ found for minimizing an easier problem with an outstanding optimizer can serve as priors for a harder problem with a mediocre optimizer. This realization of meta-learning for the harder problem as parameter initialization, even if the optimal parameters are not sufficiently close to the global minima, is our proposition.

\subsection{Genetic optimization}
To verify our hypothesis, we identify the easiest problem as a two-qubit MAX-CUT, which can be directly embedded in the minimal structure of our circuit~\eqref{eq:Hcircuit2}. As observed, BFGS is highly sensitive to parameter initialization, making it suboptimal for our problem. Conversely, the genetic algorithm has been utilized to optimize quantum circuits and algorithms, demonstrating its ability for global search at the expense of considerable computational resources. While employing the genetic algorithm for optimizing the harder problem may be too complex, it is deemed acceptable for solving the two-qubit case and deriving parameters with maximal fitness.

In Fig.~\ref{fig:pulse_meta}(a), we present the optimal control pulse $F[P^{\ast}(t)]$ for the two-qubit problem, achieving errorless ground state preparation. The corresponding parameters $\{A_i^*,\phi_i^*\}$ are then employed to initialize the BFGS optimizer for the 8-qubit problem, resulting in an initial error rate of $R=0.744$. This initialization leads to $F[\tilde{P}^*(t)]$ [see Fig.\ref{fig:pulse_meta}(b)] that converges to $R=0.056$, which is remarkably close to the exact ground state.

\begin{figure}
\includegraphics[width=\linewidth]{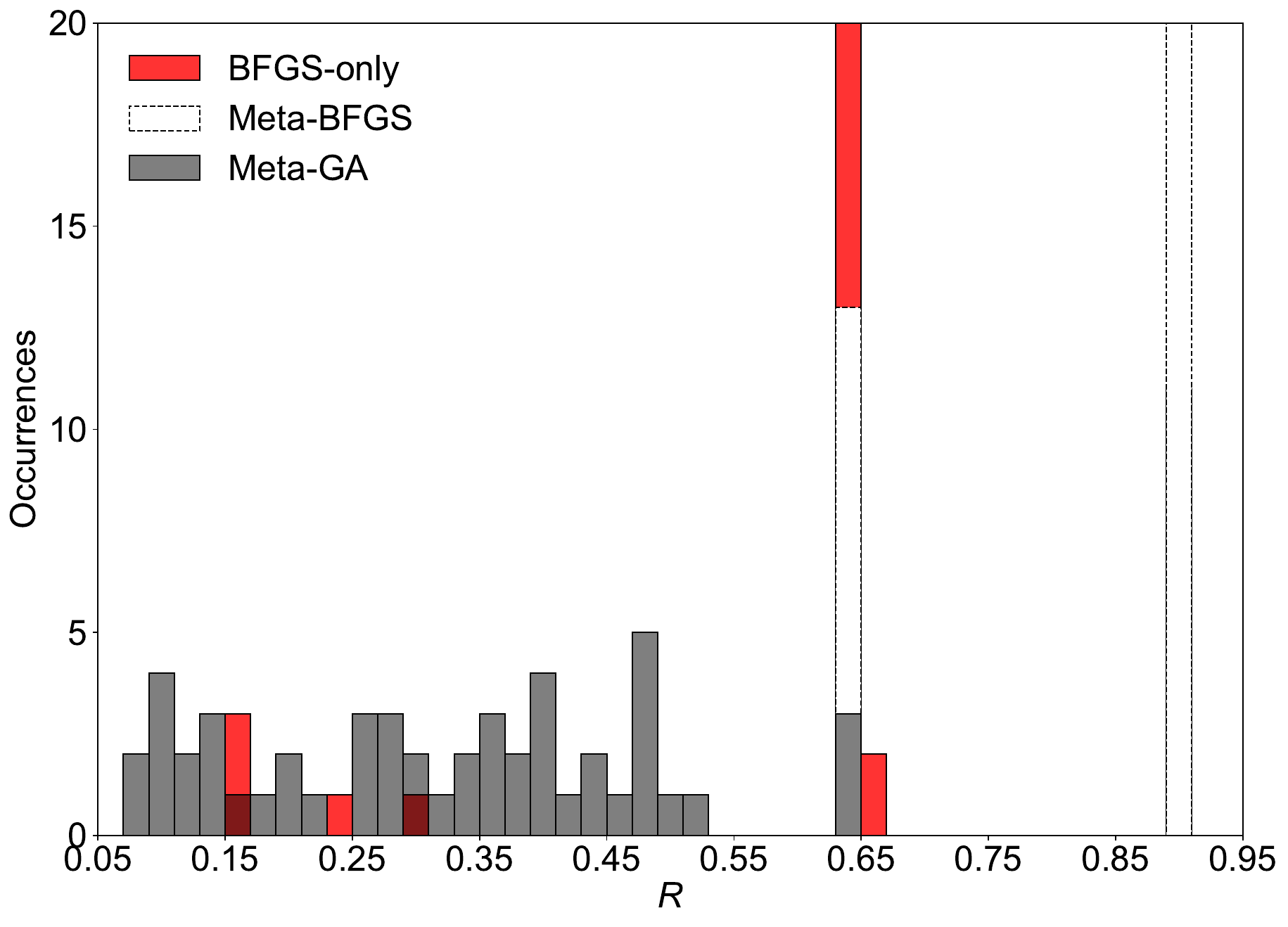}
\caption{Histogram of BFGS (baseline), Meta-learning with BFGS, and Meta-learning with genetic algorithm (GA) for 8-qubit QAOA. \label{fig:histogram}}
\end{figure}
Furthermore, we establish the statistical significance of meta-learning with the genetic algorithm over BFGS, demonstrating that the example in Fig.~\ref{fig:pulse_meta} is not cherry-picked. In Fig.~\ref{fig:histogram}, we depict the distribution of error rates $R$ for the baseline (BFGS for 8-qubit), meta-learning with BFGS, and meta-learning with the genetic algorithm via histograms. In meta-learning, we solve a two-qubit problem with the local optimizer BFGS and global optimizer GA, respectively, finding the exact solution $\langle\Psi_f|H_p|\Psi_f\rangle=-1$. \jonzen{Although both optimizers achieve good results on simpler problems, using two sets of optimal parameters as initialization for the local optimizer BFGS to solve the more challenging 8-qubit problem leads to different model performances. Therefore, we observe that meta-learning with the genetic algorithm outperforms the other two approaches.} This finding supports the notion that employing a global optimizer to optimize an easy problem and using the result as parameter initialization for a harder problem is effective. In constrast, utilizing a local optimizer for an easier problem (see Meta-BFGS) does not overcome the local minimum in the harder one, even if it also solves the easier problem well. By showcasing the superiority of meta-learning with the genetic algorithm, we provide the evidence supporting the efficacy of our approach in enhancing the performance of PBVQO.

\section{Discussion}
\label{discussion}

\jonzen{After presenting the results of the numerical experiments with PBVQO, we proceed to compare them with conventional QAOA. As discussed in Sec.~\ref{sec:hardware}, our digital-analog superconducting architecture facilitates direct implementation of XX rotation using gate Hamiltonian, avioding the need for CNOTs and local rotations. To establish a fair comparison, we initially assess the error rate of QAOA. Given that the performance of QAOA heavily depends on the depth $p$ (equivalent to optimizing $2p$ variables), we set $p=3$ for conventional QAOA, aligning with the 6 parameters in the pulse ansatz of PBVQO. Testing with $50$ sets of random initial parameters, all of them converge to $R=0.25$ for the $8$-qubit MAX-CUT problem.}

\jonzen{Another critical aspect is runtime, as longer runtimes accumulate errors due to quantum noise, particularly impactful in superconducting circuits. QAOA's runtime is directly related to single and two-qubit gate times, implemented by corresponding gate Hamiltonians. The scale of Hamiltonians dictates gate times by requiring more energetic input within a shorter operation time. Along this, we calculate the energetic cost as~\cite{Abah_2019}: 
\begin{equation}
    C=\frac{1}{T}\int_{0}^{T} ||H(t)||dt,
\end{equation}
where $||H(t)||$ represents the Frobenius norm of the total Hamiltonian of PBVQO and the quantum gates in QAOA. A higher energetic cost indicates a longer runtime, or alternatively, a more intense pulse if the operation time is bounded. After averaging $50$ independent simulations, we find $C_{\text{QAOA}}=488.76$ for QAOA, while  PBVQO incurs almost half the cost at $C_{\text{PBVQO}}=280.87$.} 

\jonzen{While demonstrating the capability of our superconducting architecture in solving quantum optimization problems with PBVQO, surpassing conventional QAOA, it's essential to note that it is far from the optimal architecture for MAX-CUT problems.} For instance, the regular QAOA embedding requires all-to-all connectivity thus, for being implemented in our platform it needs performing swapping operations that add more complexity from the point of view of compilation and pulse levels because now the pulses anzats requires learning such information. Alternatives approaches will include flux-qubit or transmon system coupled to tunable couplers~\cite{Sungprx,heunisch2023}.\par

\fc{It is important to note that this pulse-based variational approach isn't limited to superconducting circuits or circuit quantum electrodynamics architecture. The flexibility of our approach only requires a quantum platform capable of implementing time-dependent couplings and single-qubit rotations. Promosing candidates should be Rydberg atoms~\cite{Saffman2016, Levine2019, Browaeys2020} where one- or two-dimensional arrays of heavy alkali atoms such as Rubidium (Rb) and Cesium (Cs) are confined in optical twezzers, and single-qubit gates can be implemented by either microwave drives~\cite{Dotsenko2004}, optical stimulated Raman transitions or a combination of microwave drives with magnetic field gradients~\cite{Saffman2016} achieving gate fidelity for single-qubit rotations of $\mathcal{F}=99.792\%$~\cite{Wang2016}. Likewise, two-qubit interactions in Rb atoms has been implemented using the the blockade effect with individual and separate addresing~\cite{Saffman2016} and also with local spin exchange with atoms in movable tweezers~\cite{kaufman2015}.\par
Similarly, trapped-ions~\cite{Bruzewicz2019,Monroe2021} are viable for implementing our QAOA protocol. In this case, the microwave controlled single-qubit operations are performed in the hyperfine energy levels, while the optical ones are made using stimulated Raman transitions for inducing quadupolar transitions on the $D\rightarrow S$ levels~\cite{Bruzewicz2019}, these approaches have achieved single-qubit fidelities around $\mathcal{F}=99.993\%$~\cite{Harty2014} for microwave and $\mathcal{F}=99.995\%$~\cite{bermudez2017} for optical induced gates, respectively. Furthermore, there exist a plethora of ways for implementing two-qubit interaction in the platform; from Coulomb interaction~\cite{Bruzewicz2019}, using coherently the motional degree of freedom of the quantized motion~\cite{Cirac1995}, and using such degree of freedom as a mediator for implementing a multi-qubit interaction~\cite{MS1999}. Theses approaches have achieved gate fidelities of $\mathcal{F}=99.91\%$~\cite{gaebler2016} for hyperfine M\o{}lmer-S\o{}rensen gate.}



\section{Conclusion and outlook}
\label{conclusion}

In summary, we have introduced pulse-based variational quantum optimization (PBVQO) as an extension to variational quantum algorithms in the NISQ era. Our protocol is directly applicable to the superconducting circuit architecture within the digital-analog quantum computing paradigm. Through experiments solving the MAX-CUT problem, we have demonstrated PBVQO's capability to approximate the ground state of the problem Hamiltonian.
Furthermore, we have proposed meta-learning for parameter initialization in more challenging problems, utilizing a genetic algorithm as the global optimizer. The positive outcomes of meta-learning with the global optimizer have been illustrated through benchmarking against baseline PBVQO and meta-learning with a local optimizer. Additionally, we have conducted comparisons with conventional gate-based QAOA and analyzed the mechanisms underlying its advantages.

Overall, the outlook for PBVQO is promising, with numerous open questions and opportunities for future research and development. For instance, investigating new optimization algorithms or adapting existing classical optimization methods to better suit the requirements of pulse-level optimization could lead to further improvements in performance and convergence speed.  By extending pulse-based approaches to other prominent variational algorithms, such as VQE, one can potentially unlock new capabilities and address a wider range of optimization problems.
Finally, as PBVQO continues to evolve, rigorous experimental validation and benchmarking against classical and quantum algorithms are essential. Thus, future study can be dedicated to the comprehensive experimental studies for validating the scalability, robustness, and performance of pulse-based techniques on noisy quantum hardware platforms, with error migitation \cite{RMP2023noise}.

\begin{acknowledgements}
This work is supported by NSFC (12075145 and 12211540002), STCSM (2019SHZDZX01-ZX04), and the Innovation Program for Quantum Science and Technology (2021ZD0302302), HORIZON-CL4-2022-QUANTUM-01-SGA project 101113946 OpenSuperQPlus100 of the EU Flagship on Quantum Technologies, the Basque Government through Grant No. IT1470-22. F.A.C.L. thanks to the German Ministry for Education and Research, under QSolid, Grant no. 13N16149. This project has also received funding from the European Union's HORIZON Europe program via project HORIZON-CL4-2021-DIGITALEMERGING-02-10 (No. 101080085 QCFD). 
\end{acknowledgements}

\end{document}